\documentclass[10pt,twocolumn,conference]{IEEEtran}

\IEEEoverridecommandlockouts

\usepackage{graphicx}
\usepackage{amsmath}
\usepackage{amssymb}
\usepackage[caption=false]{subfig}
\usepackage[noadjust]{cite}
\usepackage{float}
\usepackage{algorithm}
\usepackage{bibentry}
\usepackage{balance}
\usepackage{algorithm}
\usepackage{algorithmicx}
\usepackage{algpseudocode}
\usepackage{xcolor}
\usepackage{subfig}
\usepackage{url}

\graphicspath{{img/}}

\begin{document}

\bstctlcite{IEEEexample:BSTcontrol}

\title{Out-of-Zone Signal Leakage Sensing in Radio Dynamic Zones\thanks{This work is supported in part by the NSF award CNS-1939334 and its associated supplement for studying National Radio Dynamic Zones (NRDZs).}}

\author{
\IEEEauthorblockN{Sung Joon Maeng, \.{I}smail G\"{u}ven\c{c}, Mihail L. Sichitiu, and Ozgur Ozdemir}\IEEEauthorblockA{Department of Electrical and Computer Engineering, North Carolina State University, Raleigh, NC\\
\{smaeng, iguvenc, mlsichit, oozdemir\}@ncsu.edu}}

\maketitle

\begin{abstract}
Radio dynamic zones (RDZs) are geographically bounded areas where novel advanced wireless technologies can be developed, tested, and improved, without the concern of interfering  to other incumbent radio technologies nearby the RDZ. In order to operate an RDZ, use of a real-time spectrum  monitoring system carries critical importance. Such a monitoring system should detect out-of-zone (OoZ) signal leakage outside of the RDZ, and if the interference to nearby receivers is intolerable, the monitoring system should be capable of mitigating such interference. This can e.g. be achieved by stopping operations inside the RDZ or switching to other bands for RDZ operation. In this paper, we introduce a spectrum monitoring concept for OoZ signal leakage detection at RDZs, where sensor nodes (SNs) are installed at the boundary of an RDZ and monitor the power leakage from multiple transmitters within the RDZ. We propose a prediction algorithm that estimates the received interference at OoZ geographical locations outside of the RDZ, using the measurements obtained at sparsely located SNs at the RDZ boundary. Using computer simulations, we evaluate the performance of the proposed algorithm and study its sensitivity to SN deployment density. 
\end{abstract}

\begin{IEEEkeywords}
    Correlated shadowing, estimation, interpolation, Kriging, national radio dynamic zone (NRDZ).  
\end{IEEEkeywords}

\section{Introduction}\label{sec:intro}
Radio dynamic zones (RDZs) are geographically bounded areas where existing receivers nearby the RDZ are protected from high power transmitters that are located inside the RDZ. The transmitters include directed energy systems, high-power microwave transmitters, and experimental systems. For example, in the United States (US), national radio dynamic zones (NRDZs)~\cite{NRDZ_vs_NRQZ} are recently being conceptualized to serve similar goals. In particular, NRDZ testbeds in the US are expected to \emph{``foster innovation by enabling new paradigms in dynamic spectrum sharing between electromagnetic transmitters and receivers''} in the future, \emph{``while enabling transition-to-practice through robust partnerships with public- and private-sector organizations''}, in close coordination with the US National Science Foundation,   Federal Communications Commission (FCC), and the National Telecommunications and Information Administration (NTIA)~\cite{NSF_NRDZ_DCL}. 

It is not easily feasible to manage and control harmful interference from individual RDZ transmitters to incumbent sensitive receivers if such transmitters are closely located to the receivers. Therefore, establishing an isolated geographical zone for introducing spatial separation from sensitive receivers,  and monitoring interference leakage outside of the geographical zone, carries critical importance. The RDZ is then required to have a prompt reaction to dynamic changes in the environment to protect the receivers. 

There are several works in the existing literature related to real-time spectrum monitoring. For example, radio emission maps (REMs) are utilized for evaluating the coverage of the cellular networks as studied in~\cite{zhao2007applying,yilmaz2013radio}. In \cite{li2009distributed,ma2008soft,thilina2013machine,jin2018privacy}, crowd-sourced and cooperative spectrum sensing have been explored. However, none of these studies consider the specific RDZ concept and they do not address the problem of accurate signal detection outside of a geographically bounded area.  
The measurements from the SNs for spectrum monitoring are commonly characterized by spatially correlated shadow fading in the literature~\cite{cressie2015statistics}. In works such as \cite{braham2016fixed,sato2017kriging}, Kriging-based power prediction and spatial interpolation of the measurements are used in the shadowing auto-correlation model. However, these studies are limited to a single transmit source scenario.
Cross-correlation models are extensively surveyed and explored \cite{szyszkowicz2010feasibility,graziosi2002general}, which enables the description of the correlation between signals from different sources.  

\begin{figure}[t]
    \centering
    \includegraphics[width=0.38\textwidth]{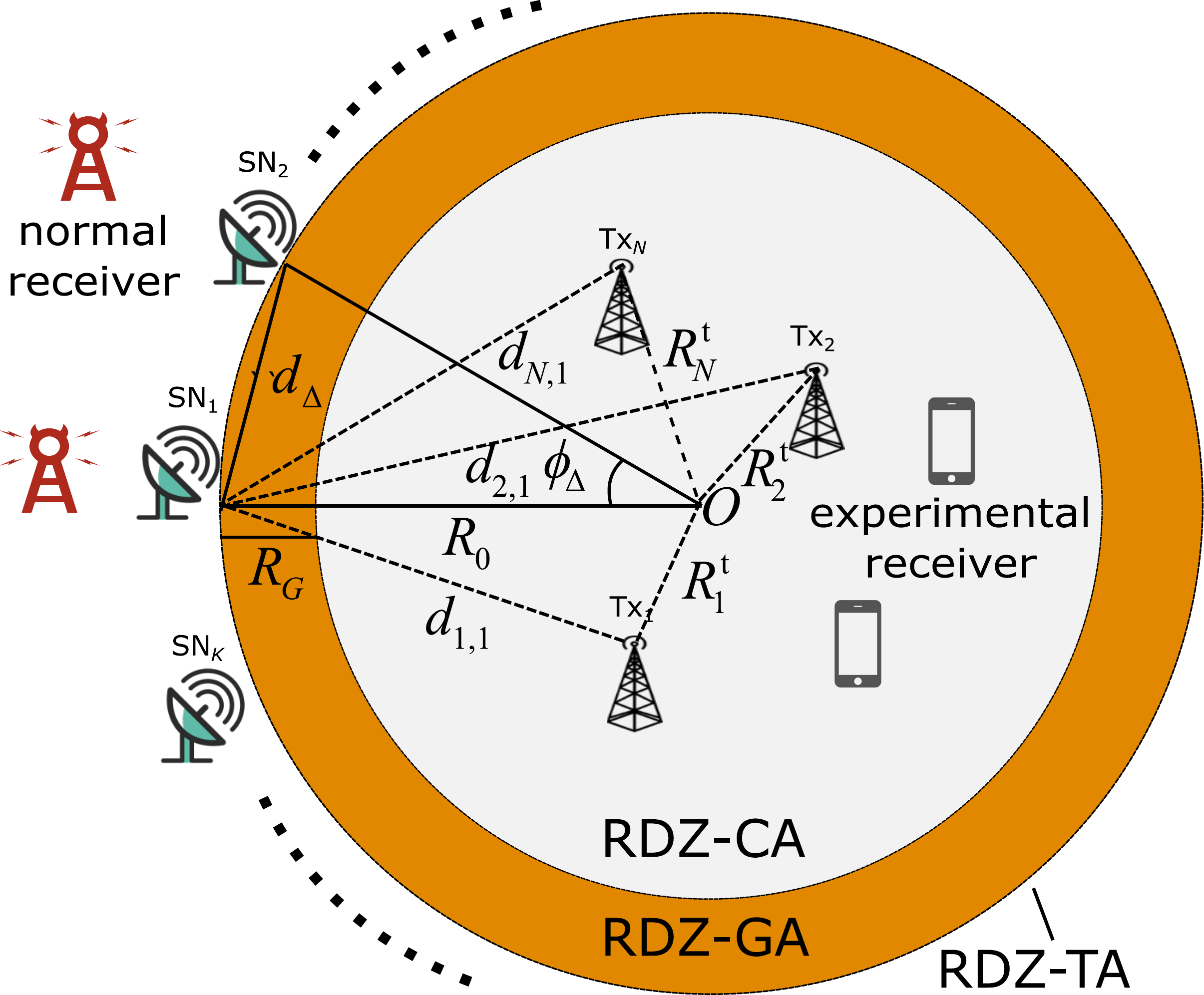}
    \caption{Illustration of the considered spectrum leakage monitoring approach for a radio dynamic zone (RDZ).}
    \label{fig:NRDZ_design}
\end{figure}

In this paper, we study how an RDZ efficiently deploys the SNs and monitors the out-of-zone (OoZ) signal leakage surrounding the area. Sensor nodes (SNs) are sparsely deployed at the boundary of the zone and they measure the received power from multiple transmitters inside the RDZ.
We consider both the cross-correlation as well as the auto-correlation of signals using a Kriging based approach to deal with signals from multiple transmitters. We propose techniques to estimate the parameters of the correlation models and we introduce a signal prediction algorithm based on spatial correlation in the RDZ scenario. Using computer simulations, we study the performance of the proposed approach and compare it with an approach where spatial correlation is not taken into account. Our results show that with a sparse SNs deployment setup, the prediction performance of our proposed algorithm is superior to the path-loss-based algorithm where the spatial correlation is not considered. In addition, the performance of proposed algorithm improves as the radius of the RDZ ($R_0$) and angle spacing ($\theta_{\Delta}$) becomes smaller.

\section{System Model}\label{sec:system}
\subsection{Proposed RDZ Concept}

In this subsection, we introduce our proposed concept for an RDZ, that includes transmitters, receivers, and monitoring SNs as in Fig.~\ref{fig:NRDZ_design}. RDZ seeks to protect the normal receivers outside of the RDZ, for example, cell phones, from the special transmitters inside of the RDZ. 
There are $N$ transmitters that are considered to be deployed in the RDZ core area (RDZ-CA), but are restricted  to be deployed in the RDZ guard area (RDZ-GA) to introduce a physical isolation between the RDZ transmitters and OoZ receivers. The fixed $K$ SNs are located at the boundary of RDZ target area (RDZ-TA). They measure and predict the power leakage from the RDZ to establish the real-time radio environment boundary map. 

The location of the transmitters and the SNs are represented by the distance and the azimuth angle from the origin as
\begin{align}
    l^{\rm t}_{n}&=({\rm R}_{n}^{\rm t},\phi^{\rm t}_{n}),\;l_{k}^{\rm r}=({\rm R}_{0},\phi_{k}^{\rm r}),
\end{align}
where the distance and the azimuth angle of the transmitters are uniformly distributed as ${\rm R}_{n}^{\rm t}\sim\mathcal{U}[0,{\rm R}_{0}-{\rm R}_{\rm G}]$, $\phi^{\rm t}_{n}\sim\mathcal{U}[0,2\pi]$, and the SNs surround the boundary of the zone with uniform angle spacing ($\phi_{\Delta}$) as $\phi^{\rm r}_{k}=(k-1)\phi_{\Delta}$. The width of the RDZ-GA is given by ${\rm R}_{\rm G}$. It is worth noting that the angle spacing should be small enough to guarantee the prediction of the signal power at OoZ locations, while it should be large enough  to keep the SN infrastructure cost low. Furthermore, we can decide the width of the guard area (${\rm R}_{\rm G}$) depending on the signal power level from the transmitters in order to manage the OoZ signal leakage. The distance between two adjacent SNs can be written as
\begin{align}\label{eq:d_del}
    d_{\Delta}&={\rm R}_{0}\sqrt{2(1-\cos(\phi_{\Delta})}\approx{\rm R}_{0}\phi_{\Delta}.
\end{align}
In addition, the distance between $n_{\rm th}$ transmitter and  $k_{\rm th}$ SN can be expressed as
\begin{align}
    d_{n,k}&=\|l^{\rm t}_{n}-l^{\rm r}_{k}\|\nonumber\\
    &=\sqrt{({\rm R}_{n}^{t})^2+({\rm R}_{0})^2-2{\rm R}_{n}^{t}{\rm R}_{0}\cos(|\phi^{\rm t}_{n}-\phi^{\rm r}_{k}|)},
\end{align}
which can be obtained by the laws of cosines.

\subsection{Radio Propagation Model}

\begin{figure}[t]
    \centering
    \includegraphics[width=0.2\textwidth]{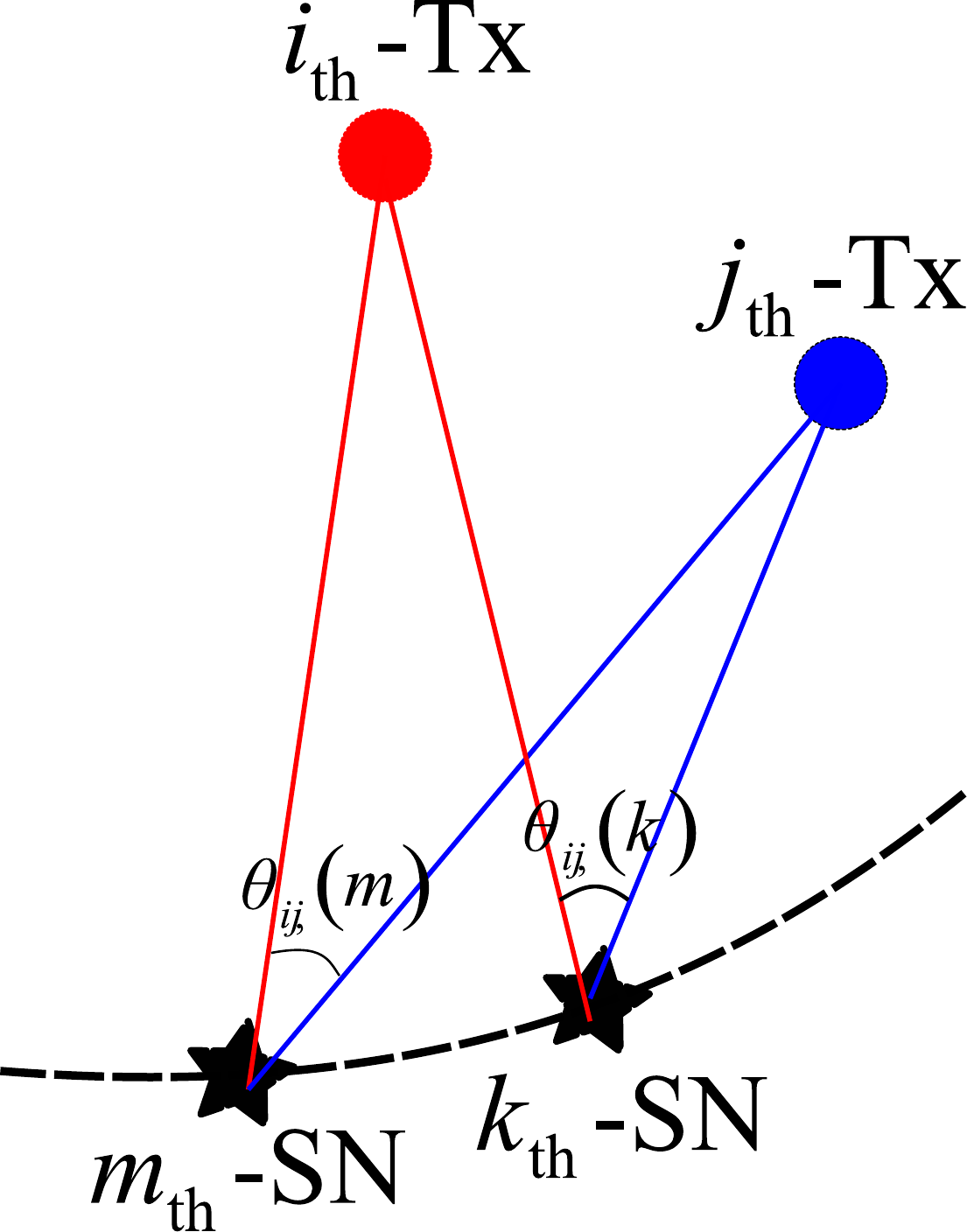}
    \caption{Correlation model for transmitted and received signals with two transmitters and two SNs.}
    \label{fig:corr_model}
\end{figure}
The received signal power can be described by the attenuation of the transmit power and the fluctuation of the power from the different environmental characteristics, and can be expressed as~\cite{graziosi1999multicell}:
\begin{align}\label{eq:rec_sig}
    y_{n}(k)&=\mathsf{P}_{\rm Tx}-10\eta\log_{10}(d_{n,k})+w_{n}(k),
\end{align}
where $\mathsf{P}_{\rm Tx}$, $\eta$ denote the transmit power and the path-loss coefficient, and $w_{n}(k)$ indicates the shadow fading component that is modeled by a zero-mean Gaussian process with a spatial covariance function. In particular, $w_{n}(k)$ captures the correlation between two signals at different locations. It can also be represented by jointly lognormal shadowing where $w_{1}(k),\cdots,w_{N}(k)$ are jointly Gaussian. An SN can receive signals from multiple transmitters simultaneously, and we assume that the SN is able to observe signals from different sources separately. This assumption can be feasible by periodic pilot transmissions from the transmitters with coordination between transmitters and the SNs in centralized networks. Then, the received signal vector at an SN that stacks the signals from all transmitters witin the RDZ can be expressed as
\begin{align}
    \textbf{y}(k)&=\left[y_1(k),\dots,y_N(k)\right]^{\rm T}.
\end{align}

\subsection{Correlation Model for Shadow Fading}

In this subsection, we describe a correlation model between the received signals of SNs from multiple transmitters, as illustrated in Fig.~\ref{fig:corr_model}. Four received signals in Fig.~\ref{fig:corr_model} need to be characterized by individual correlation models. For example, the correlation between received signals of two SNs from a source can be modeled by a auto-correlation function (ACF), while the correlation between received signals from the different sources can be modeled by a cross-correlation function (CCF). Besides, the CCF needs to be divided into two separate models depending on whether two received signals are coming from the same transmitter or not.

\subsubsection{Auto-Correlation Model}

The ACF between the received signals at $k_{\rm th}$ and $m_{\rm th}$ SNs, for the transmission from the $i_{\rm th}$ transmitter, can be expressed as
\begin{align}\label{eq:auto_corr}
    \rho_{i,i}(k,m)&=\frac{\mathbb{E}\left[w_{i}(k)w_{i}(m)\right]}{\sigma_{w}^2}=\exp{\left(-\frac{\|l^{\rm r}_k-l^{\rm r}_m\|}{d_{\rm cor}}\ln{2}\right)},
\end{align}
where $\sigma_{w}$, $d_{\rm cor}$ denote the standard deviation of shadow fading component and the correlation distance. We adopt the model that the correlation exponentially decays as the distance between the SNs become larger~\cite{sato2017kriging}. We assume a homogeneous environment where the shadowing log-variance ($\sigma_{w}^2$) is constant regardless of distance and sources. Then, the auto-correlation depends only on the distance between the SNs, and it is independent of the transmitter, given by
\begin{align}
    \rho_{i,i}(k,m)&=\rho_{j,j}(k,m).
\end{align}

\subsubsection{Cross-Correlation Model}
Unlike the auto-correlation function, it is commonly known that the cross-correlation of shadowing is related to the angle difference between two sources. The cross-correlation between two shadowing components of the $k_{\rm th}$ SN from the $i_{\rm th}$, $j_{\rm th}$ transmitter can be expressed as 
\begin{align}\label{eq:cross-corr}
    \rho_{i,j}(k,k)&=\frac{\mathbb{E}\left[w_{i}(k)w_{j}(k)\right]}{\sigma_{w}^2}=A\cos(\theta_{i,j}(k))+B,
\end{align}
where $A+B\le1$~\cite{graziosi1999multicell}. The angle $\theta_{i,j}(k)$ can be derived by the law of cosine as follow:
\begin{align}
    \cos(\theta_{i,j}(k))&=\frac{(d_{i,k})^2+ (d_{j,k})^2-\|l^{\rm t}_i-l^{\rm t}_j\|^2}{2d_{i,k}d_{j,k}}.
\end{align}
Next, we obtain the cross-correlation between the remaining pair of shadowing components, signals from different sources, and different SNs. The cross-correlation between two shadowing components of the $k_{\rm th}$, $m_{\rm th}$ SNs from the $i_{\rm th}$, $j_{\rm th}$ transmitter can be expressed as
\begin{align}\label{eq:cross_cor_2}
    \frac{\mathbb{E}\left[w_{i}(k)w_{j}(m)\right]}{\sigma_{w}^2}=\rho_{i,j}(k,m)\overset{(a)}{=}\rho_{i,i}(k,m)\rho_{i,j}(k,k)\nonumber\\
    =\left(A\cos(\theta_{i,j}(k))+B\right)\exp{\left(-\frac{\|l^{\rm r}_k-l^{\rm r}_m\|}{d_{\rm cor}}\ln{2}\right)},
\end{align}
where (a) comes form \cite[(3)]{graziosi2002general}. Note that jointly stationary assumption in (a) holds if $\theta_{i,j}(k)\approx\theta_{i,j}(m)$. It implies that this model should take into account of the sufficiently small variation of the angle in designing parameter $\theta_{\Delta}$. Furthermore, $d_{\Delta}$ should be decided by considering $d_{\rm cor}$. For example, if $d_{\Delta}$ is relatively too large compared with $d_{\rm cor}$, the correlation between two adjacent SNs becomes really low, which results in poor prediction performance.

\section{RDZ Model Parameter Estimation}

In this section, we describe how an RDZ can estimate unknown parameters related to path-loss and shadow fading that are introduced in the previous section, using the measurements obtained at the SNs.

\subsection{Estimation of the Path-loss Coefficient ($\eta$)}

The received signal power in \eqref{eq:rec_sig} can be rewritten as
\begin{align}
    y_{n}(k)&=\textbf{t}_n(k)\boldsymbol{\beta}+w_n(k),\nonumber\\
    \textbf{t}_n(k)&=\left[\mathsf{P}_{\rm Tx}\quad -10\log\left(d_{n,k}\right)\right],~\boldsymbol{\beta}=\left[1\quad \eta\right]^{\rm T}.
\end{align}
We assume that the RDZ is aware of the locations of the transmitters and their transmit power $\mathsf{P}_{\rm Tx}$. The ordinary least square (OLS) solution to estimate $\eta$ can be formulated as
\begin{align}\label{eq:OLS}
    \hat{\boldsymbol{\eta}}&=\arg\min\sum_{k}\sum_{n}\left[y_n(k)-\textbf{t}_n(k)\boldsymbol{\beta}\right]^2.
\end{align}
The OLS solution minimizes the square error of all measurements of SNs (total $K\times N$). Equivalently, it is given by
\begin{align}
    \hat{\boldsymbol{\beta}}&=\left(\textbf{T}^{\rm T}\textbf{T}\right)^{-1}\textbf{T}^{\rm T}\textbf{Y},
\end{align}
where $\textbf{T}=\left[\textbf{t}_1(1);\dots;\textbf{t}_N(K)\right]$, $\textbf{Y}=\left[y_1(1),\dots,y_N(K)\right]^{\rm T}$. After we estimate path-loss, we can extract the shadowing component from the received signal power as follows:
\begin{align}\label{eq:w_est}
    \hat{w}_n(k)=y_n(k)-\textbf{t}_n(k)\hat{\boldsymbol{\beta}}.
\end{align}

\subsection{Estimation of the Cross-Correlation Function Coefficient}

In this subsection, cross-correlation coefficients in \eqref{eq:cross-corr} are estimated by a RDZ. The shadow fading components from N transmitters at the $k_{\rm th}$ SN follow multivariate Gaussian distribution, which is given by
\begin{align}
    \textbf{w}(k)\sim N(0,\textbf{C}(k,k)),
\end{align}
where $\textbf{C}(k,k)\in\mathbb{R}^{N\times N}$ is the covariance matrix, and ($i_{\rm th},j_{\rm th}$) entry of the matrix is given as
\begin{align}\label{eq:cosine_model}
    \textbf{C}_{i,j}(k,k)=\rho_{i,j}(k,k)=A\cos(\theta_{i,j}(k))+B,
\end{align}
where each entry is the same as \eqref{eq:cross-corr}. If the mean and the covariance matrix are given, the log-likelihood function can be written as
\begin{align}
    L(\textbf{w}(k))&=-\frac{1}{2}\log|\textbf{C}(k,k)|-\frac{1}{2}\textbf{w}(k)^{\rm T}\textbf{C}(k,k)^{-1}\textbf{w}(k)+C,
\end{align}
where $C$ is a constant value. Then, we can estimate coefficients $A$ and $B$ using maximum likelihood (ML) estimation as:
\begin{align}\label{eq:MLE}
    &(\hat{A},~\hat{B})=\nonumber\\
    &\arg\max\sum_{k=1}^{K}\left(-\frac{1}{2}\log|\textbf{C}(k,k)|-\frac{1}{2}\textbf{w}(k)^{\rm T}\textbf{C}(k,k)^{-1}\textbf{w}(k)\right).
\end{align}

\subsection{Estimation of the Auto-Correlation Function Coefficient}

In this section, we estimate the ACF in \eqref{eq:auto_corr}. The spatial dependence of shadowing can be modeled with a semivariogram, which is given by:
\begin{align}
    \gamma(k,m)=\frac{1}{2}\text{var}\left(y(k)-y(m)\right).
\end{align}
We can empirically estimate semivariogram by using the method of moments. We utilize the fact that SNs are uniformly deployed so that the distance between any two SNs is an integer multiple of $d_{\Delta}$. For example, the distance between the two adjacent SNs is $d_{\Delta}$, while the distance between the SN and the second next SN is $2d_{\Delta}$. The estimated semivariogram can then be expressed as
\begin{align}\label{eq:semi_var_est}
    \hat{\gamma}(Md_{\Delta})&=\frac{1}{2N(K-M)}\sum_{n=1}^{N}\sum_{k=1}^{K-M}\left(y_n(k+M)-y_n(k)\right)^2.
\end{align}
By using the fact that covariance in \eqref{eq:auto_corr} is modeled by an exponential form, the semivariogram can be modeled as
\begin{align}
    \gamma(Md_{\Delta})=D\left(1-\exp{\left(-\frac{Md_{\Delta}}{E}\right)}\right)~,
\end{align}
where $D$, $E$ are the coefficients that we try to find. Then, these two coefficients can be estimated by the least square (LS) estimator as follows:
\begin{align}\label{eq:LSE}
    (\hat{D},\hat{E})&=\arg\min\left(\hat{\gamma}(Md_{\Delta})-\gamma(Md_{\Delta};D,E)\right)^2.
\end{align}
We can find $\hat{d}_{\rm cor}$ from $\hat{E}$ by \eqref{eq:auto_corr} and the following relation between covariance and semivariogram~\cite[2.3.23]{cressie2015statistics}:
\begin{align}
    \gamma(k,m)=\sigma_{w}^2-\sigma_{w}^2\rho(k,m).
\end{align}
Note that since the total number of measurements ($N(K-M)$) decreases as $M$ increases in \eqref{eq:semi_var_est}, and that fewer measurements can degrade the accuracy of estimation, we only consider $M=1,~2,~3$ in simulations.

\section{Interpolation to OoZ Locations Using Kriging}

In this section, using the measurements from all the available SNs and using an estimated correlation model, we interpolate sensing measurements to other OoZ locations where there are no SNs located. 

\subsection{Ordinary Kriging Method}
The ordinary Kriging method minimizes the spatial prediction error at an  arbitrary receiver location with the linear combination of the measurements data set. The ordinary Kriging can formulate the problem statement as follows:
\begin{IEEEeqnarray}{rl}
    \min_{\lambda_{1,1},\dots,\lambda_{N,K_{\rm s}}}
    &\quad \mathbb{E}\left[\left(\hat{w}_n(k_0)-w_n(k_0)\right)^2\right],\\
    \text{s.t.}
    &\quad \hat{w}_n(k_0)=\sum_{i=1}^{N}\sum_{k=1}^{K_{\rm s}}\lambda_{i,k}w_i(k),\\
    &\quad \sum_{i=1}^{N}\sum_{k=1}^{K_{\rm s}}\lambda_{i,k}=1 \IEEEyessubnumber,
\end{IEEEeqnarray}
where $k_0$ is a location with no SN measurements. Note that we do not utilize all measurement data, but local data from a few close distance SNs where $\textbf{K}_{\rm s}\in \textbf{K}$ indicates measurements from the local SNs. 

To solve the above problem, we need to transform it to an equivalent problem by using Lagrange multipliers as follows~\cite{cressie2015statistics}:
\begin{align}\label{eq:prob_Lagr}
    \min_{\lambda_{1,1},\dots,\lambda_{N,K_{\rm s}}}&\mathbb{E}\left[\left(w_n(k_0)-\sum_{i=1}^{N}\sum_{k=1}^{K_{\rm s}}\lambda_{i,k}w_i(k)\right)^2\right]\nonumber\\
    &-\mu\left(\sum_{i=1}^{N}\sum_{k=1}^{K_{\rm s}}\lambda_{i,k}-1\right),
\end{align}
where $\mu$ denotes Lagrange multiplier. The objective function in \eqref{eq:prob_Lagr} can be rewritten as
\begin{align}\label{eq:obj_rewri}
    &\sigma^2_{w}+\sigma^2_{w}\sum_{j=1}^{N}\sum_{m=1}^{K_{\rm s}}\sum_{i=1}^{N}\sum_{k=1}^{K_{\rm s}}\lambda_{j,m}\lambda_{i,k}\textbf{C}_{i,j}(k,m)\nonumber\\
    &-2\sigma^2_{w}\sum_{i=1}^{N}\sum_{k=1}^{K_{\rm s}}\lambda_{i,k}\textbf{C}_{n,i}(k_{0},k)-\mu\left(\sum_{i=1}^{N}\sum_{k=1}^{K_{\rm s}}\lambda_{i,k}-1\right)
\end{align}
where $\textbf{C}_{i,j}(k,m)=\mathbb{E}\left[w_{i}(k)w_{j}(m)\right]/\sigma_{w}^2=\rho_{i,j}(k,m)$. Next, we can find the solution by obtaining the first derivative of \eqref{eq:obj_rewri} and equating the result to zero, which is given by
\begin{align}\label{eq:first_der}
    &\sum_{j=1}^{N}\sum_{m=1}^{K_{\rm s}}\lambda_{j,m}\textbf{C}_{i,j}(k,m)-\textbf{C}_{n,i}(k_{0},k)+\mu'=0,\nonumber\\
    &\quad i=1,\dots,N,~k=1,\dots,K_{\rm s}.
\end{align}
We can also formulate \eqref{eq:first_der} as follows:
\begin{align}\label{eq:matrix_form}
    &\left[ \begin{array}{c c c c}
        \textbf{C}_{1,1}(1,1) & \cdots & \textbf{C}_{1,N}(1,K_{\rm s}) & 1\\
        \textbf{C}_{2,1}(1,1) & \cdots & \textbf{C}_{2,N}(1,K_{\rm s}) & 1\\
        \vdots & \vdots & \vdots & \vdots\\
        \textbf{C}_{N,1}(K_{\rm s},1) & \cdots & \textbf{C}_{N,N}(K_{\rm s},K_{\rm s}) & 1\\
        1 & \cdots & 1 & 0\\
    \end{array}\right]
    \left[ \begin{array}{c}
        \lambda_{1,1}\\\lambda_{2,1}\\ \vdots\\ \lambda_{N,K_{\rm s}}\\ \mu'
    \end{array}\right]\nonumber\\
    &=\left[ \begin{array}{c}
        \textbf{C}_{n,1}(k_0,1)\\\textbf{C}_{n,2}(k_0,1)\\ \vdots\\ \textbf{C}_{n,N}(k_0,K_{\rm s})\\1
    \end{array}\right].
\end{align}
We can then obtain optimal $\lambda_{1,1},\dots,\lambda_{N,K_{\rm s}}$ from \eqref{eq:matrix_form} and estimate the received signal powers at the target locations of interest as follows:
\begin{align}\label{eq:Kriging_sol}
    \hat{w}_n(k_0)=\sum_{i=1}^{N}\sum_{k=1}^{K_{\rm s}}\lambda_{i,k}w_i(k).
\end{align}

The proposed estimation and  interpolation algorithms are summarized in Algorithm 1. If an RDZ does not support the proposed Kriging interpolation in Algorithm 1, it is able to estimate the signal power by using path-loss based estimation as follows:
\begin{align}\label{eq:pathloss_est}
    \hat{y}_n(k_0)=\mathsf{P}_{\rm Tx}-10\hat{\eta}\log_{10}(d_{n,k_0}).
\end{align}
Since the path-loss-based estimation does not take into account spatial correlation, the accuracy would be worse than Kriging interpolation.

\begin{algorithm}
  \caption{RDZ Kriging interpolation}
  \label{algorithm:main}
    \begin{algorithmic}[1]
        \State \textbf{Initialize:} $\textbf{y}(1),\dots,\textbf{y}(K)$, and the location information of transmitters and SNs.
        \State Estimate $\hat{\eta}$ by OLS solution in \eqref{eq:OLS}.
        \State Estimate $\hat{w}_1(1),\dots,\hat{w}_N(K)$ by \eqref{eq:w_est}.
        \State Compute $\hat{A}$, $\hat{B}$ by ML estimation in \eqref{eq:MLE}.
        \State Estimate semivariogram $\hat{\gamma}(Md_{\Delta})$ by \eqref{eq:semi_var_est}.
        \State Obtain $\hat{D},\hat{E}$ by LS estimator in \eqref{eq:LSE}.
        \For{all desired points $k_0$}
        \State Obtain $\hat{w}_n(k_0)$ by ordinary Kriging in \eqref{eq:matrix_form} and \eqref{eq:Kriging_sol}.
        \State Compute received signal power by $\hat{y}_n(k_0)=\mathsf{P}_{\rm Tx}-10\hat{\eta}\log_{10}(d_{n,k_0})+\hat{w}_n(k_0)$.
        \EndFor
    \end{algorithmic}
\end{algorithm}

\subsection{Generation of Jointly Lognormal Shadowing Components}

In our shadow fading model, $\textbf{w}=[w_n(1),\dots,w_N(K)]^{\rm T}$ are jointly Gaussian with spatial covariance matrix $\textbf{C}$. When $\textbf{C}$ is a symmetric positive semi-definite matrix (\textit{psd}), it can be factorized as 
\begin{align}
    \textbf{C}=\textbf{L}\textbf{L}^{T}.
\end{align}
The covariance matrix from the exponential model in \eqref{eq:auto_corr} and the cosine model in \eqref{eq:cosine_model} is known as \textit{psd}, and the multiplication of the two separable models is \textit{psd} as well~\cite{szyszkowicz2010feasibility}. If $\textbf{C}$ is a positive definite matrix, $\textbf{L}$ can be generally obtained by Cholesky decomposition. Then, we generate independent Gaussian random variables $\textbf{z}\sim \mathcal{N}(0,\sigma_{\rm w}^2\textbf{I})$ and obtain shadow fading components as follows:
\begin{align}
    \textbf{w}=\textbf{L}\textbf{z}.
\end{align}

When we try to generate $\textbf{w}$ in our scenario and interpolate signal power at OoZ locations by Algorithm 1, 
it is not feasible to generate $\textbf{w}$ of whole OoZ locations by considering the correlation between every location. First, the size of covariance matrix $\textbf{C}$ would be too large to compute factorization and multiplication. For example, the size of covariance matrix is $\textbf{C}\in\mathbb{R}^{(N \times  (l^{\rm r}_{\rm gri}+K))\times(N \times  (l^{\rm r}_{\rm gri}+K))}$ where $l^{\rm r}_{\rm gri}$ indicates the number of all points in the gird of desired locations to predict. Second, the jointly stationary assumption in \eqref{eq:cross_cor_2} does not hold if the spatial distance between two points is too large. In this case, by using the marginal distributions property, we can locally generate $\textbf{w}$ by just generating the subset of the random variables $\textbf{w}_{\rm sub}\subset\textbf{w}$ and dropping the irrelevant variables from the covariance matrix. Therefore, when we predict the received power at an OoZ location, we can build a covariance matrix that only includes the local SNs measurement data that is used to predict with the OoZ location that we are interested in.

\begin{table}[!t]
\renewcommand{\arraystretch}{1.1}
\caption{System parameters settings}
\label{table:settings}
\centering
\begin{tabular}{lc}
\hline
Parameter & Value \\
\hline\hline
Transmit power ($\mathsf{P}_\mathsf{Tx}$) & 30 dBm\\
Radius of RDZ-TA ($R_0$) & [500 1000] m\\
Width of RDZ-GA ($R_{\rm G}$) & $R_0/10$ m\\
Angle spacing ($\phi_{\Delta}$) & [5 10 15 20 30] $^{\circ}$\\
Number of transmitters (N) & 3 \\
Number of SNs (K) & $360/\phi_{\Delta}$\\
path-loss coefficient ($\eta$) & 3.5\\
Standard deviation of shadow fading ($\sigma_{w}$) & 8 dB\\
Correlation distance ($d_{\rm cor}$) & 100 m\\
Cross-correlation coefficients (A, B) & [$0.7$ $0.3$]\\
\hline
\hline
\end{tabular}
\vspace{-0.15in}
\end{table}
\begin{figure}[t]
    \centering
    \includegraphics[width=0.50\textwidth]{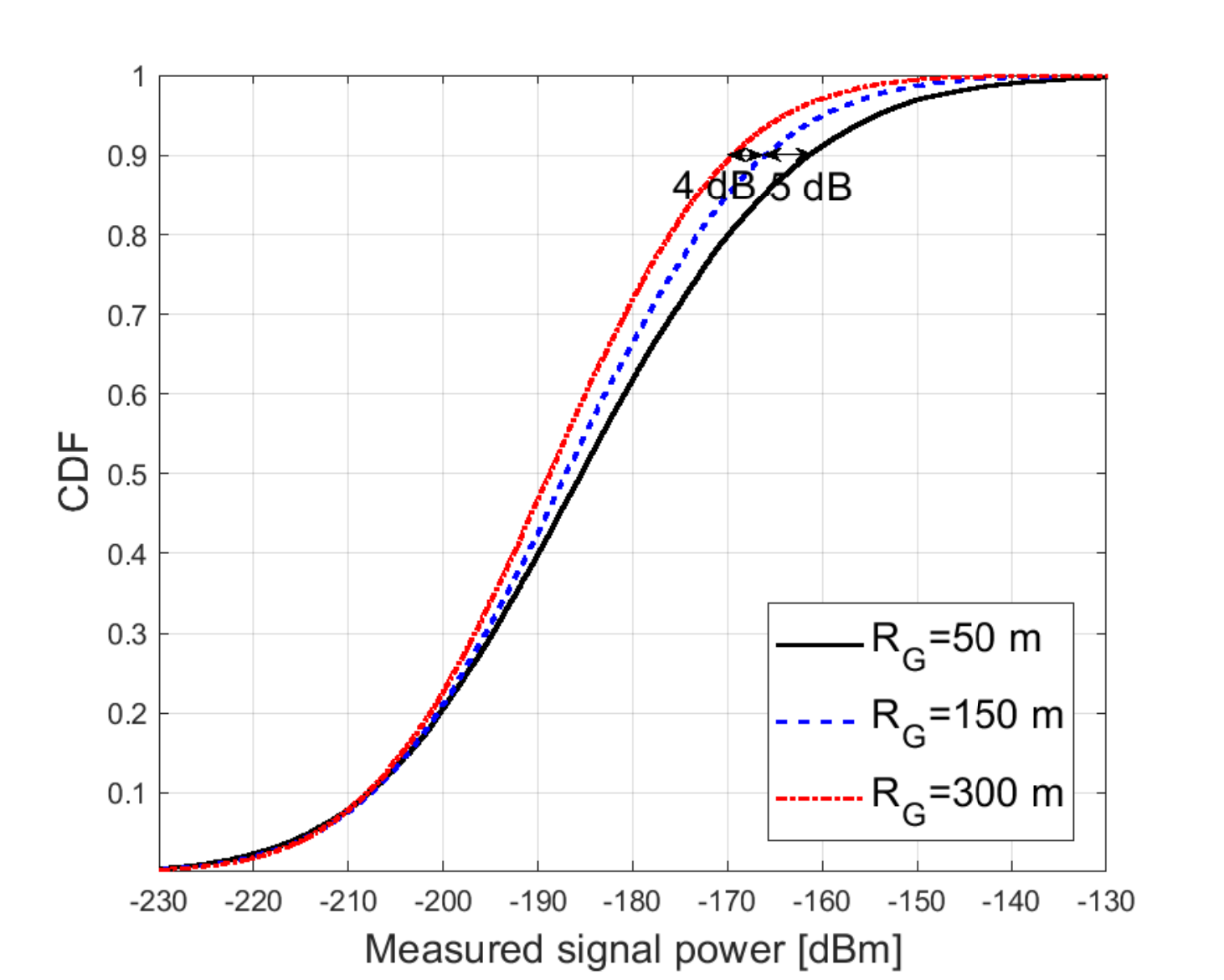}
    \caption{The CDF of the measured signal power at SNs depends on $R_{\rm G}$.}
    \label{fig:Power_RG}
\end{figure}
\section{Numerical Results}\label{sec:simulation}

\begin{figure}[t]
    \centering
    \includegraphics[width=0.50\textwidth]{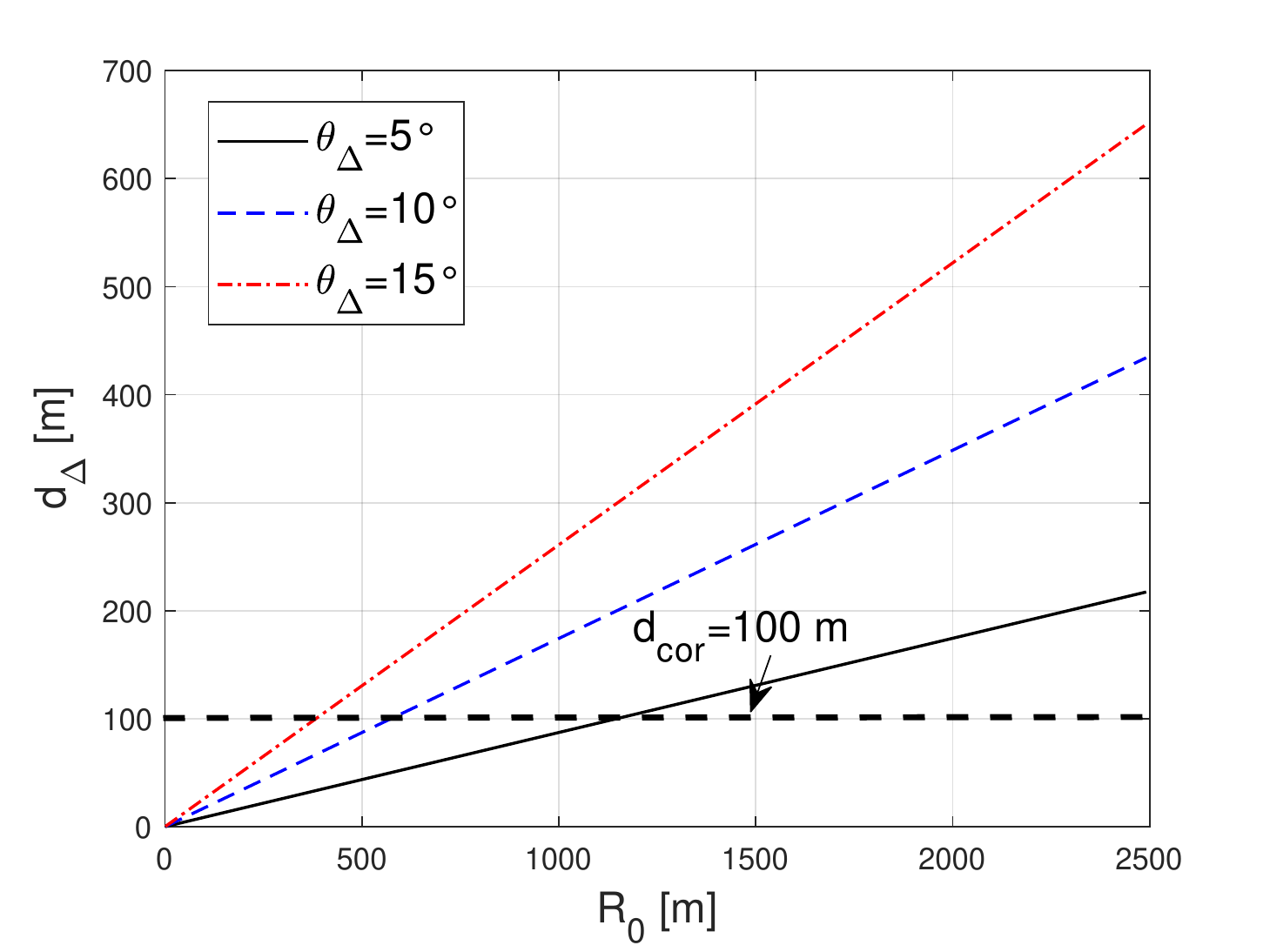}
    \caption{The increase of $d_{\Delta}$ as $R_0$ grows}
    \label{fig:d_del}
\end{figure}

In this section, we evaluate the performance of the proposed algorithm considering various different parameter configurations and scenarios. The system parameters are listed in Table \ref{table:settings}. In Fig,~\ref{fig:Power_RG}, we show the change of the distribution of measured signal power at the SNs depending on the width of RDZ-GA ($R_{\rm G}$). It is observed that larger $R_{\rm G}$ is effective in reducing the signal power at the boundary of a RDZ since the average distance between transmitters and the boundary increases. The measured power gaps at 90 percent are 5 dB and 4 dB, as $R_{\rm G}$ grows from 50 m to 150 m and 300 m.
Fig.~\ref{fig:d_del}, on the other hand, shows how the  distance between adjacent SNs ($d_{\Delta}$) grows as a function of $R_0$ and $\theta_{\Delta}$ as captured in \eqref{eq:d_del}. By \eqref{eq:auto_corr}, the correlation between measurements of adjacent SNs decreases as $d_{\Delta}$ grows and when it is the same as $d_{\rm cor}$, the correlation goes to 0.5. As the correlation becomes lower, the estimation performance from the Kriging interpolation becomes poor.

\begin{figure}[t]
    \centering
    \includegraphics[width=0.50\textwidth]{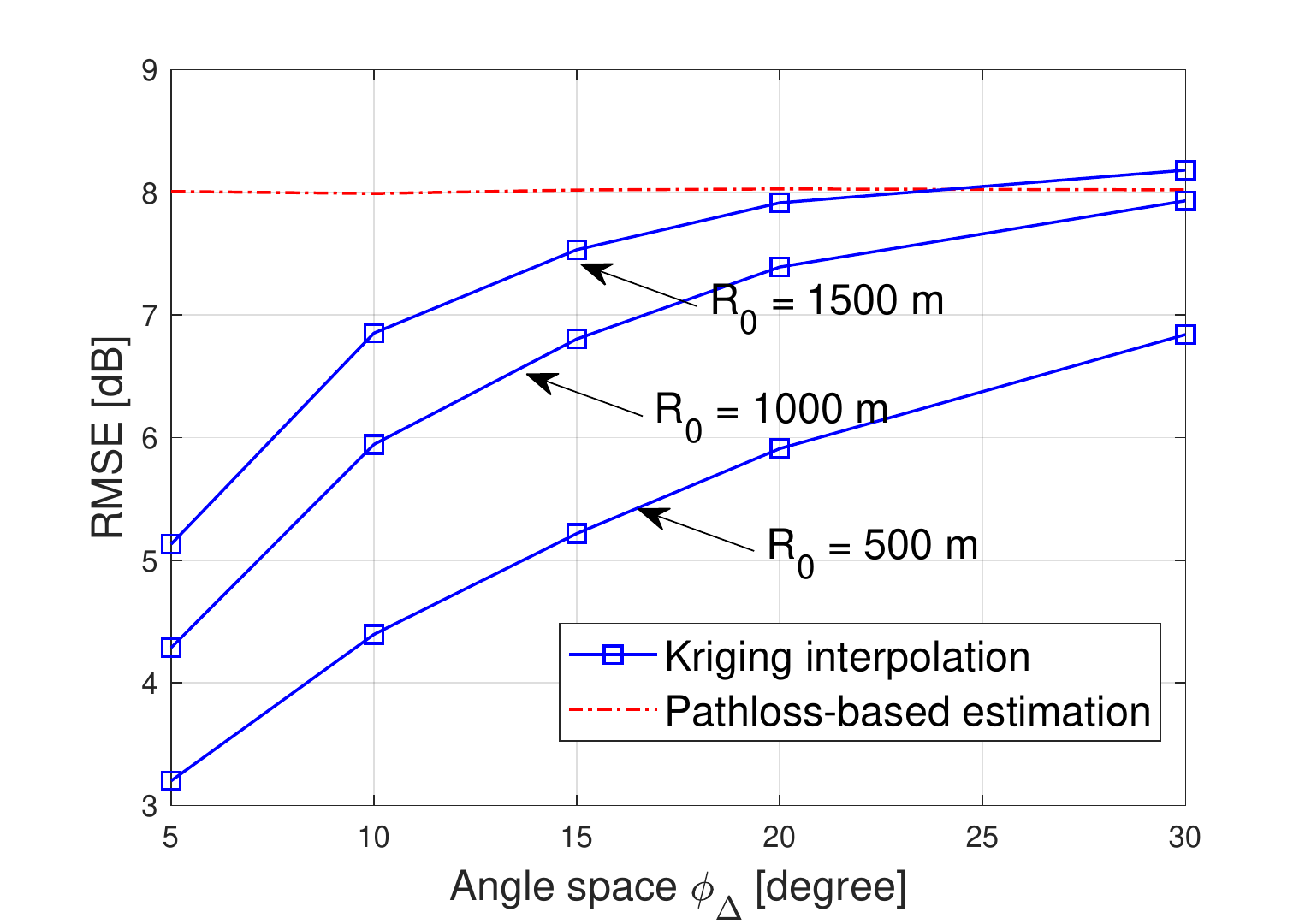}
    \caption{The RMSE of the prediction by the proposed Kriging interpolation algorithm depending on $R_0$ and $\theta_{\Delta}$.}
    \vspace{-0.2in}
    \label{fig:interp_RMSE}
\end{figure}

The performance of the proposed Kriging interpolation in Algorithm 1 is shown in Fig.~\ref{fig:interp_RMSE}. The path-loss-based estimation result is obtained by \eqref{eq:pathloss_est}. We iteratively calculate the signal power of an unknown location and the predicted signal power by Algorithm 1 and compare them with the root mean square error (RMSE). The RMSE is calculated by $\sqrt{\frac{1}{N_{\rm iter}}\sum^{N_{\rm iter}}\left(y_n(k_0)-\hat{y}_n(k_0)\right)^2}$. Since path-loss-based estimation does not take into account the spatial correlation and does not interpolate signals from SNs, the RMSE is constant regardless of the angle spacing. On the other hand, the Kriging interpolation algorithm improves the RMSE as the angle spacing and the radius of the zone become smaller. Although the Kriging interpolation outperforms the path-loss-based estimation in most cases, the performance could be worse than the path-loss-based estimation if the angle spacing and the radius of the zone are too large.

\begin{figure}[t]
    \centering
    \subfloat[$\eta$, $d_{\rm cor}$]{
    \includegraphics[width=0.50\textwidth]{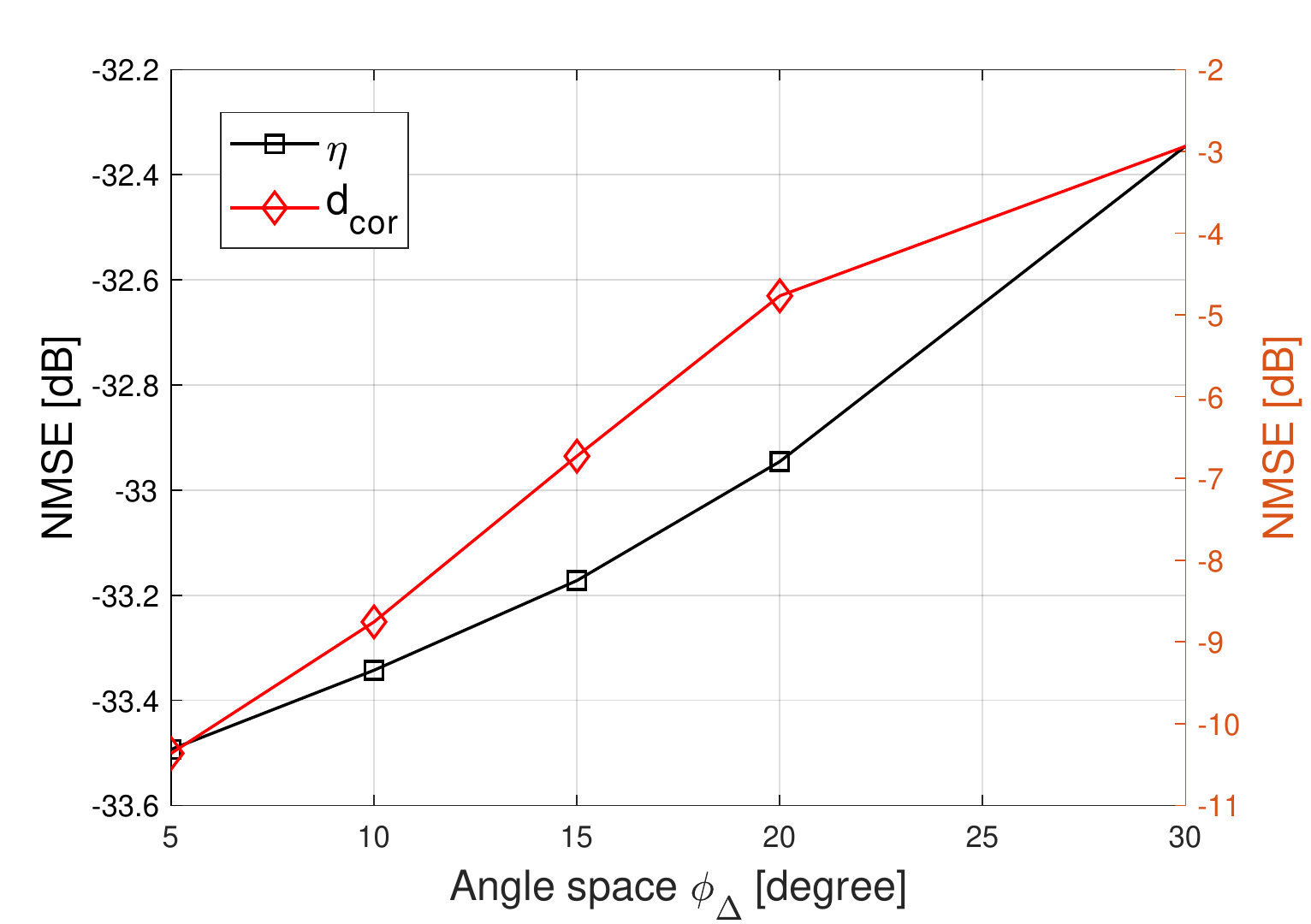}}
    
    \subfloat[A, B]{
    \includegraphics[width=0.50\textwidth]{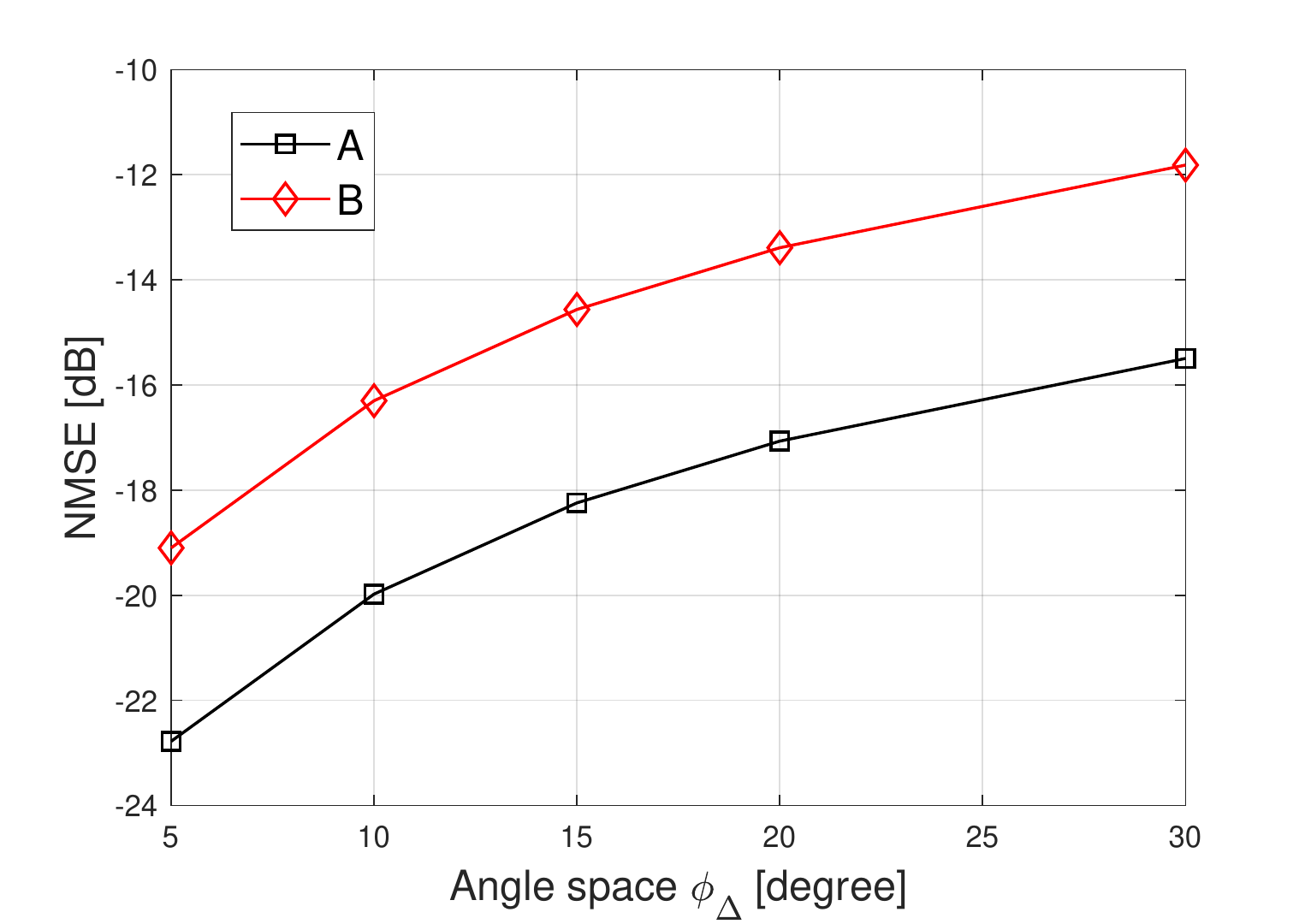}}
    \caption{The NMSE of coefficients estimations where $R_0=500$ m.}
    \vspace{-0.2in}
    \label{fig:NMSE_par}
\end{figure}

Fig.~\ref{fig:NMSE_par} shows the accuracy of the path-loss and correlation coefficients estimation using the normalized mean square error (NMSE) depending on the angle spacing. The NMSE of $\eta$ can be calculated by $\sqrt{\frac{1}{N_{\rm iter}}\sum^{N_{\rm iter}}\left(\frac{\eta-\hat{\eta}}{\eta}\right)^2}$. It is observed that the NMSE of all coefficients degrades as angle spacing increases. This is due to the fact that the number of SNs that are used in estimating the coefficients is reduced.

\section{Conclusion}\label{sec:conclusion}
An RDZ is a geographical area that has special rules for enabling novel spectrum experiments, and the power leakage escaping from the RDZ should be controlled to protect normal receivers outside the zone. In this paper we introduce an RDZ concept and a power leakage monitoring algorithm with spatially sparse installment of SNs around the RDZ. Using simulations, we characterize the performance of the proposed estimation and the prediction algorithms and highlight associated trade-offs between the accuracy of the prediction technique and the density of the SNs.  

\vspace{-0.1in}

\bibliographystyle{IEEEtran} 
\bibliography{IEEEabrv,bibfile}

\begin{thebibliography}{10}
\providecommand{\url}[1]{#1}
\csname url@samestyle\endcsname
\providecommand{\newblock}{\relax}
\providecommand{\bibinfo}[2]{#2}
\providecommand{\BIBentrySTDinterwordspacing}{\spaceskip=0pt\relax}
\providecommand{\BIBentryALTinterwordstretchfactor}{4}
\providecommand{\BIBentryALTinterwordspacing}{\spaceskip=\fontdimen2\font plus
\BIBentryALTinterwordstretchfactor\fontdimen3\font minus
  \fontdimen4\font\relax}
\providecommand{\BIBforeignlanguage}[2]{{%
\expandafter\ifx\csname l@#1\endcsname\relax
\typeout{** WARNING: IEEEtran.bst: No hyphenation pattern has been}%
\typeout{** loaded for the language `#1'. Using the pattern for}%
\typeout{** the default language instead.}%
\else
\language=\csname l@#1\endcsname
\fi
#2}}
\providecommand{\BIBdecl}{\relax}
\BIBdecl

\bibitem{NRDZ_vs_NRQZ}
\BIBentryALTinterwordspacing
T.~Kidd, ``National radio quiet and dynamic zones,'' CHIPS -- The Department of
  Navy's Information Technology Magazine, Apr.-June 2018. [Online]. Available:
  \url{https://www.doncio.navy.mil/CHIPS/ArticleDetails.aspx?ID=10299}
\BIBentrySTDinterwordspacing

\bibitem{NSF_NRDZ_DCL}
\BIBentryALTinterwordspacing
``{Dear Colleague Letter: Supplemental Funding Opportunity to explore
  feasibility of National Radio Dynamic Zones (NRDZ)},'' NSF 20-079, May 2020.
  [Online]. Available:
  \url{https://www.nsf.gov/pubs/2020/nsf20079/nsf20079.jsp}
\BIBentrySTDinterwordspacing

\bibitem{zhao2007applying}
Y.~Zhao, L.~Morales, J.~Gaeddert, K.~K. Bae, J.-S. Um, and J.~H. Reed,
  ``Applying radio environment maps to cognitive wireless regional area
  networks,'' in \emph{Proc. IEEE Int. Symp. New Frontiers Dynamic Spectrum
  Access Networks}, Dublin, Ireland, 2007, pp. 115--118.

\bibitem{yilmaz2013radio}
H.~B. Yilmaz, T.~Tugcu, F.~Alag{\"o}z, and S.~Bayhan, ``Radio environment map
  as enabler for practical cognitive radio networks,'' \emph{IEEE Commun.
  Mag.}, vol.~51, no.~12, pp. 162--169, Dec. 2013.

\bibitem{li2009distributed}
Z.~Li, F.~R. Yu, and M.~Huang, ``A distributed consensus-based cooperative
  spectrum-sensing scheme in cognitive radios,'' \emph{IEEE Trans. Veh.
  Technol.}, vol.~59, no.~1, pp. 383--393, Sep. 2009.

\bibitem{ma2008soft}
J.~Ma, G.~Zhao, and Y.~Li, ``Soft combination and detection for cooperative
  spectrum sensing in cognitive radio networks,'' \emph{IEEE Trans. Wireless
  Commun.}, vol.~7, no.~11, pp. 4502--4507, Dec. 2008.

\bibitem{thilina2013machine}
K.~M. Thilina, K.~W. Choi, N.~Saquib, and E.~Hossain, ``Machine learning
  techniques for cooperative spectrum sensing in cognitive radio networks,''
  \emph{IEEE J. Sel Areas Commun.}, vol.~31, no.~11, pp. 2209--2221, 2013.

\bibitem{jin2018privacy}
X.~Jin and Y.~Zhang, ``Privacy-preserving crowdsourced spectrum sensing,''
  \emph{IEEE/ACM Trans. Netw.}, vol.~26, no.~3, pp. 1236--1249, 2018.

\bibitem{cressie2015statistics}
N.~Cressie, \emph{Statistics for spatial data}.\hskip 1em plus 0.5em minus
  0.4em\relax John Wiley \& Sons, 2015.

\bibitem{braham2016fixed}
H.~Braham, S.~B. Jemaa, G.~Fort, E.~Moulines, and B.~Sayrac, ``Fixed rank
  kriging for cellular coverage analysis,'' \emph{IEEE Trans. Veh. Technol.},
  vol.~66, no.~5, pp. 4212--4222, Aug. 2016.

\bibitem{sato2017kriging}
K.~Sato and T.~Fujii, ``Kriging-based interference power constraint: Integrated
  design of the radio environment map and transmission power,'' \emph{IEEE
  Trans. Cogn. Commun. Netw.}, vol.~3, no.~1, pp. 13--25, Jan. 2017.

\bibitem{szyszkowicz2010feasibility}
S.~S. Szyszkowicz, H.~Yanikomeroglu, and J.~S. Thompson, ``On the feasibility
  of wireless shadowing correlation models,'' \emph{IEEE Trans. Veh. Technol.},
  vol.~59, no.~9, pp. 4222--4236, Sep. 2010.

\bibitem{graziosi2002general}
F.~Graziosi and F.~Santucci, ``A general correlation model for shadow fading in
  mobile radio systems,'' \emph{IEEE Communications Letters}, vol.~6, no.~3,
  pp. 102--104, Aug. 2002.

\bibitem{graziosi1999multicell}
F.~Graziosi, M.~Pratesi, M.~Ruggieri, and F.~Santucci, ``A multicell model of
  handover initiation in mobile cellular networks,'' \emph{IEEE Trans. Veh.
  Technol.}, vol.~48, no.~3, pp. 802--814, May 1999.

\end{thebibliography}

\end{document}